# Increasing Parallelism in the ROOT I/O Subsystem

G Amadio[1], B P Bockelman[2], P Canal[3], D Piparo[1], E Tejedor[1], Z Zhang[2]

[1]CERN, [2]University of Nebraska-Lincoln, [3]FNAL

E-mail: `amadio@cern.ch`

**Abstract.** When processing large amounts of data, the rate at which reading and writing can take place is a critical factor. High energy physics data processing relying on ROOT is no exception. The recent parallelisation of LHC experiments' software frameworks and the analysis of the ever increasing amount of collision data collected by experiments further emphasised this issue underlying the need of increasing the implicit parallelism expressed within the ROOT I/O.

In this contribution we highlight the improvements of the ROOT I/O subsystem which targeted a satisfactory scaling behaviour in a multithreaded context. The effect of parallelism on the individual steps which are chained by ROOT to read and write data, namely (de)compression, (de)serialisation, access to storage backend, are discussed. Performance measurements are discussed through real life examples coming from CMS production workflows on traditional server platforms and highly parallel architectures such as Intel Xeon Phi.

## 1. Introduction

Particles collide at the Large Hadron Collider (LHC) approximately half a billion times per second. Each collision creates particles that decay into a complex cascade of many other particles that are recorded by the detectors as a series of electronic signals that are then digitally reconstructed and stored for further analysis. At the current rate, physicists have about 30 petabytes of new data per year to analyse [2], and this rate is expected to increase ten-fold in the next decade. Most of this data is stored and processed in ROOT's file format [1], hence optimizing ROOT to read and write all this data as fast as possible is of critical importance to the future of high energy physics.

*1.1. The Meaning of ROOT I/O*
One of the key innovations of ROOT is its columnar data format. With ROOT, if only a small subset of the data is needed for a given analysis, only those parts of the data file will be read, instead of the whole file. Another important aspect of ROOT I/O is its ability to perform serialisation and deserialisation of generic C++ objects via automatically generated object streamers. Moreover, ROOT can compress data before writing it to disk and decompress it before processing. Hence, ROOT I/O is much more than simply reading and writing from a regular file; it involves also (de)serialisation and (de)compression of data. Naturally, each of these processing phases can be parallelised independently. Since this form of parallelism is usually performed in a way that is invisible to the user, it is referred to as implicit parallelism, or implicit multi-threading (IMT) in ROOT. Support for implicit parallelism has been introduced in ROOT 6.06. Since then, many improvements on this front have been added to newer versions, and this is what we discuss in the following sections.

## 2. Reading ROOT Data in Parallel

Each data set created in the ROOT file format may be read many times and by different users in their analyses, hence data reading performance is arguably the most important aspect to be considered when parallelising ROOT I/O and choosing what compression algorithm to use to store data intended for physicists to use.

*2.1. Reading Data Columns*

ROOT uses a columnar data format, and experiments' data files usually contain many different columns that contain the type, energy, momentum, and other properties of particles digitally reconstructed from collision events. One way in which it is possible to parallelise data reading is decompressing and deserialising each column in parallel. This implicit multi-threading feature has been introduced in ROOT 6.08. As a demonstration of the potential performance gains, two benchmarks have been run to read, decompress and deserialise data sets from CMS (GenSim) and ATLAS (xAOD), with approximately 70 and 200 different columns, respectively. Figure 1 shows that a speedup of about 3.5 can be obtained on a quad core laptop, a great improvement over running the code sequentially.

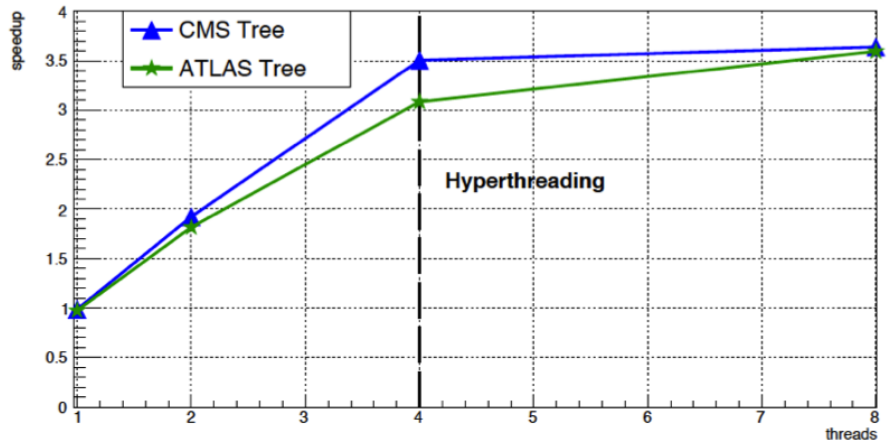

**Figure 1.** Parallel reading of multiple data columns.

*2.2. Data Decompression*

ROOT files usually store data in several compressed blocks, also referred to as baskets. When processing large files, many blocks can be decompressed and deserialised in parallel. Since decompression is compute-intensive, it is also possible to interleave parallel decompression with processing of decompressed data. This feature, currently under development, is planned for inclusion in ROOT 6.14. Figure 2 shows performance gains for the current implementation on a quad-core laptop (4 threads + hyperthreading).

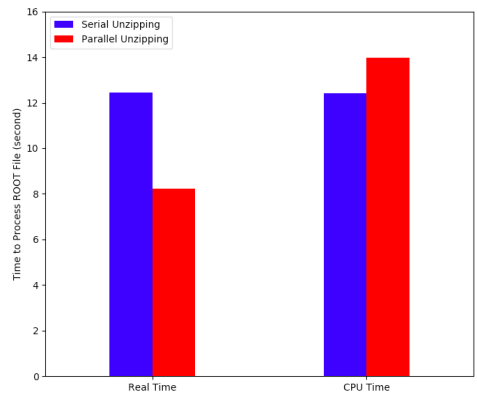

**Figure 2.** Parallel basket decompression.

## 3. Writing ROOT Data in Parallel

Even if reading ROOT data is much more common than writing, that is not to say that writing performance is not important for ROOT. Event reconstruction jobs process vast amounts of data from the detectors, so if the speed of writing out reconstructed data becomes a bottleneck, it may reduce the overall percentage of events that can be saved for offline analysis. In addition to allowing users to use one `TFile` per thread in their programs, ROOT has added new classes and features to support parallel writing, which are discussed next.

### 3.1. Writing Data Columns

In a similar fashion as what has been done for reading data columns from a file in parallel, ROOT can also serialise and compress multiple data columns in parallel when implicit multi-threading is enabled. This feature has become available in ROOT 6.08. The performance advantages of this parallelism have been demonstrated with a benchmark within CMSSW [3] that writes out two kinds of data sets: an I/O intensive reconstruction data set (RECO), and a data set with lighter I/O requirements, intended for analysis (AOD). Figure 3 shows that on the Intel Xeon Phi codenamed Knights Landing (KNL) [4], with IMT disabled, the throughput begins to degrade for the RECO data set after about 20 CMSSW streams (30 threads). When IMT is enabled, however, the throughput can be more than doubled and scales better to a larger number of threads. The number of streams is the number of framework tasks running in parallel; the number of threads is fixed to 1.5 times the number of streams. For AOD—a slimmer kind of data set—it is also possible to see an improvement, but since it is a less I/O intensive workload the gains are limited, as the performance for this data set is already close to that of not writing out any data at the end (red line in the figure).

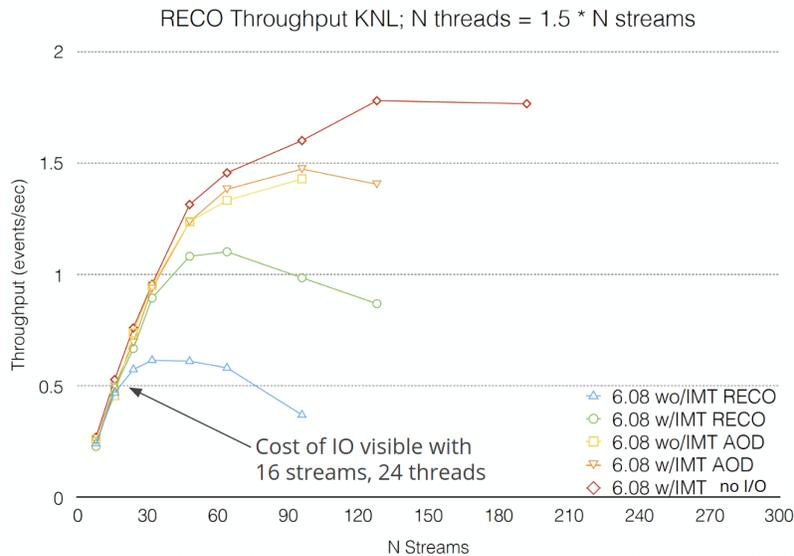

**Figure 3.** Parallel writing of multiple data columns.

### 3.2. Writing to an Output File from Multiple Threads

Since ROOT 6.06, it has been possible to write out data in parallel to multiple ROOT files, one per each thread of the user application. However, this creates the inconvenience of having many files with partial data to analyse at the end, or having an extra step to merge all data into a single file. For the user, it is much more convenient to write out data in parallel from multiple threads into a single output file by performing the merging steps automatically in the background. This feature has been introduced in ROOT 6.10 with the addition of the `TBufferMerger` [5] class. A schematic of how it works is shown below in figure 4.

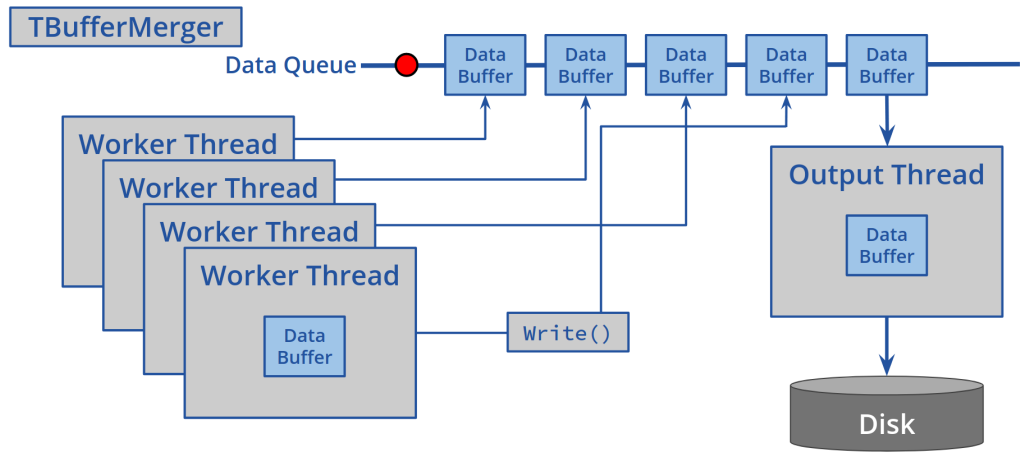

**Figure 4.** `TBufferMerger` schematic diagram. Worker threads place data in a queue, while the output thread merges one by one each buffer from the queue into the output file.

Users create a `TBufferMerger` [5] instead of a `TFile` [6], and get from it a `TMemFile`-like buffer to write into by calling `TBufferMerger::GetFile()`. Then, each worker thread can write to that `TMemFile`-like buffer, and the effect is that data will be pushed into a processing queue for merging into the output file. Once all data has been pushed into the queue and merged, the output thread closes the output file and exits. The code samples below show a simple example of filling a `TTree` in serial mode, and how that changes when using the `TBufferMerger`.

**Sequential usage of `TFile`**

```cpp
void Fill(TTree &tree, int init, int count)
{
    int n = 0;

    tree->Branch("n", &n, "n/I");

    for (int i = 0; i < count; ++i) {
        n = init + i;
        tree.Fill();
    }
}

int WriteTree(size_t nEntries)
{
    TFile f("myfile.root");
    TTree t("mytree","mytree");

    Fill(&t, 0, nEntries);

    t.Write();

    return 0;
}
```

**Parallel usage of `TFile` with `TBufferMerger`**

```cpp
void Fill(TTree *t, int init, int count);

int WriteTree(size_t nEntries, size_t nWorkers)
{
    size_t nEntriesPerWorker = nEntries/nWorkers;

    ROOT::EnableThreadSafety();
    ROOT::Experimental::TBufferMerger merger("myfile.root");

    std::vector<std::thread> workers;

    auto workItem = [&](int i) {
        auto f = merger.GetFile();
        TTree t("mytree", "mytree");
        Fill(t, i * nEntriesPerWorker, nEntriesPerWorker);
        f->Write(); // Send remaining content over the wire
    };

    for (size_t i = 0; i < nWorkers; ++i)
        workers.emplace_back(workItem,i);

    for (auto&& worker : workers) worker.join();

    return 0;
}
```

**Figure 5.** `TBufferMerger` code example.

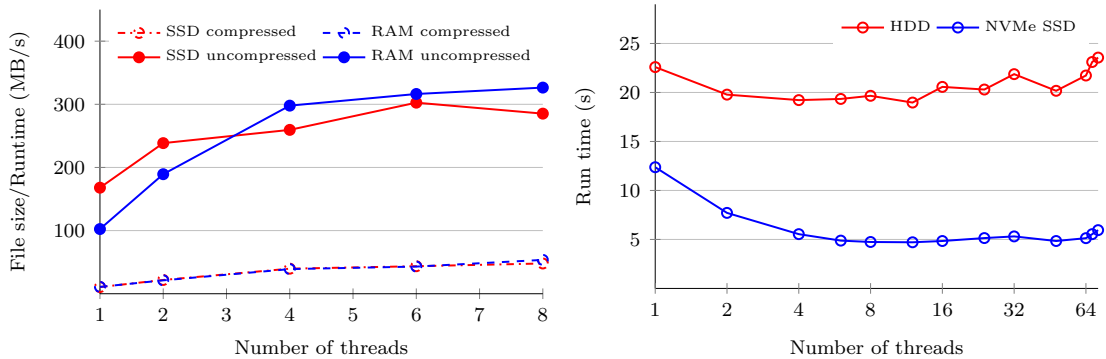

**Figure 6.** `TBufferMerger` writing performance to a hard-disk drive (HDD), a solid state drive (SSD), and a non-volatile memory solid state drive (NVMe SSD).

Write performance is shown in figure 6 for different output scenarios. The benchmarks consist in generating 1GB of pseudo-random numbers, and writing them out as a single column data file. On the left, red lines represent writing to a solid state disk (SSD), while blue lines represent writing data directly into system memory via tmpfs. On the right, two extremes in terms of hardware performance are compared: a standard hard drive (HDD) and a fast non-volatile memory solid state disk (NVMe SSD). Writing speed reaches over 320 MB/s on the SSD, which is near the hardware limit for uncompressed data. Speedup compared with sequential code is over three-fold on a quad core machine. When writing out compressed data, the CPU becomes the bottleneck due to the cost of compression. Hence, the speedup is similar, but writing speed decreases to about 50 MB/s. In this case, it is possible to scale to a larger number of threads until the limit of the disk is reached. The speedup depends on the workload and on the performance of the output device. On the right of figure 6, where only compressed data is written out, the fast NVMe SSD yields four times faster performance than a standard HDD.

### 3.3. Concurrency Improvements

Other concurrency improvements within ROOT were made possible by the development of the features just presented, such as reducing the usage of mutex locks in ROOT's interpreter where it is possible. After moving to a recent version of ROOT with these optimizations and using the `TBufferMerger` class, visible performance improvements have been achieved in CMSSW's output module [7].

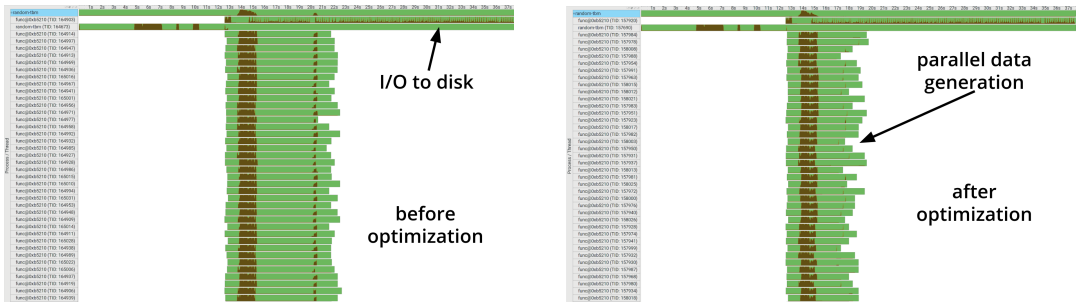

**Figure 7.** Concurrency optimisations in ROOT I/O.

In order to illustrate the effect of these optimizations, figure 7 shows VTune [8] screenshots of two runs of the benchmark on the right of figure 6, which were run on a 36-core Intel Xeon server, one before the optimisations (left), and another after (right). Each stripe represents one thread of execution (36 threads in total for this benchmark). Brown regions denote that the CPU is doing useful work and green regions denote that a thread is currently running, but not doing useful work. The single-threaded region at the beginning represents ROOT's startup, the short threads represent data generation, and the long thread at the end is the output thread writing to disk. Data generation threads are shorter after the optimisations due to the reduction in the usage of mutex locks. However, from figure 7 it is clear that what affects total running time the most is the single-threaded startup phase and the speed of the disk.

*3.4. Parallel Merging of Files with `hadd` Tool*
Finally, ROOT has incorporated several improvements to perform I/O in parallel, but users may still have old data in multiple files that could be processed and merged into a single file for further analysis. Since version 6.10, ROOT provides an option to the `hadd` tool for merging files in parallel.

## 4. Conclusion
Implicit parallelism has been introduced in ROOT beginning with version 6.06. Since then, many improvements have already been made to make ROOT I/O more performant. However, much work still lies ahead of us, if we want to cope with the expected ten-fold increase in luminosity at the LHC in the near future—which inevitably means similar increases in data production rate. For instance, in ROOT 7 old interfaces will be reformed, and in ROOT 6 new features are being added, such as more compression algorithms. Further optimizations are also planned in ROOT's interpreter and libraries. Nevertheless, current benchmarks show that ROOT is already not a bad performer when I/O is considered. Other work [9] has demonstrated that ROOT fares quite well if not even leads in performance when compared with alternatives like HDF5, protobuf, Parquet, etc, all that while offering many features not yet available in these alternative formats. ROOT is hence in the right path to continue to support the largest physics experiments in the next decade and beyond.

**References**
[1] Brun R and Rademakers F ROOT Object Oriented Data Analysis Framework *New computing techniques in physics research V. Proceedings, 5th International Workshop, AIHENP '96, Lausanne, Switzerland, September 2-6, 1996*
[2] CERN Computing Page https://home.cern/about/computing
[3] CMSSW Framework https://github.com/cms-sw/cmssw
[4] Sodani A Knights landing (KNL): 2nd Generation Intel® Xeon Phi$^{TM}$ processor *2015 IEEE Hot Chips 27 Symposium (HCS)* pp 1–24 URL http://ieeexplore.ieee.org/document/7477467
[5] TBufferMerger https://root.cern.ch/doc/master/classROOT_1_1Experimental_1_1TBufferMerger.html
[6] TFile https://root.cern.ch/doc/master/classTFile.html
[7] Riley D CMS Update *ROOT I/O Workshop, Oct 2017* URL https://indico.fnal.gov/event/15154
[8] Intel® VTune$^{TM}$ Amplifier XE https://software.intel.com/en-us/intel-vtune-amplifier-xe
[9] Blomer J A quantitative review of data formats for HEP analyses *18th International Workshop on Advanced Computing and Analysis Techniques in Physics Research (ACAT 2017)* URL https://indico.cern.ch/event/567550/contributions/2628878